\documentclass[12pt]{article}
\begin{document}

\title{{\bf Ab Initio Estimates of the Size of the Observable Universe}}

\author{
Don N. Page
\thanks{Internet address:
profdonpage@gmail.com}
\\
Department of Physics\\
4-183 CCIS\\
University of Alberta\\
Edmonton, Alberta T6G 2E1\\
Canada
}

\date{2011 July 30}

\maketitle
\large
\begin{abstract}
\baselineskip 25 pt

When one combines multiverse predictions by Bousso, Hall, and Nomura for
the observed age and size of the universe in terms of the proton and
electron charge and masses with anthropic predictions of Carter, Carr,
and Rees for these masses in terms of the charge, one gets that the age
of the universe should be roughly the inverse 64th power, and the
cosmological constant should be around the 128th power, of the proton
charge.  Combining these with a further renormalization group argument
gives a single approximate equation for the proton charge, with no
continuous adjustable or observed parameters, and with a solution that
is within 8\% of the observed value.  Using this solution gives large
logarithms for the age and size of the universe and for the cosmological
constant that agree with the observed values within 17\%.

\end{abstract}

\normalsize

\baselineskip 20.4 pt

\newpage

A goal of physics is to predict as much as possible about the universe. 
(Here I mean `predict' in the sense of deducing from theories and
assumptions about the universe, whether or not the result of the
prediction has been known by observation temporally before the
prediction is made.)  One part of this goal would be to predict the
observed constants of physics, such as the mass and charge of the proton
and of the electron, and the cosmological constant.  Second, one might
also like to predict cosmological parameters, such as the age of the
universe at which the cosmological constant or dark energy dominates
over other forms of energy density, say $t_\Lambda$.  Third, one may
desire to predict properties of us as observers, such as the time at
which we as observers exist in the universe, say $t_{\mathrm obs}$. 
Even without a complete prediction of all of these quantities, one can
make progress if one can predict relations between them, such as the
observed coincidence that $t_{\mathrm obs} \sim t_\Lambda$.  (If one
defines $t_\Lambda$ to be the time at which the age of the universe is
the inverse of the Hubble expansion rate, then this agrees with the
present age to within the small observational uncertainty, though the
surprising precision of this agreement seems likely to be just an
accident.)  The more relations that can be predicted between the
constants of physics and the parameters of the universe, the more we
understand about them.

Here I wish to note that there have been enough approximate relations
predicted between the mass and charge of the proton and electron, the
cosmological constant, and the age of the universe (e.g, the age
$t_{\mathrm obs}$ at which observers exist) that one can predict the
absolute values of all of these quantities (in Planck units) from purely
mathematical equations, using no input from observed parameters that are
other than integers (such as the number of generations of quarks and
leptons, and the number of dimensions of space, which are not yet
predicted by these arguments).  Because constants like the cosmological
constant are more than a hundred orders of magnitude away from Planck
values, it is far too much to expect the approximate relations to give
predictions with small relative errors for the quantities themselves,
but for the logarithms the predictions are moderately close to the
observed values, within 17\% or one part in six.

The fondest hopes of many physicists would be to find a theory that
predicts all the constants of physics precisely, and perhaps also the
cosmological parameters.  One might expect that observers would exist
for some range of times within the universe and so not expect absolutely
precise predictions for $t_{\mathrm obs}$.  One may or may not expect
the same for the cosmological parameters, such as $t_\Lambda$ defined
above as the time at which $Ht = 1$, since this depends not only on the
cosmological constant that is usually regarded as a constant of physics,
but also on the relation between the matter density and the spatial
curvature of the universe, which except for a spatially flat universe is
a cosmological parameter that is usually regarded as more of an initial
condition than a constant of physics that appears in the dynamical laws.

For a time it was hoped that superstring/M theory would be a predictive
theory of this type, ultimately leading to precise predictions of all
the constants of physics (since superstring/M theory has no fundamental
adjustable dimensionless constants for the dynamical theory, in
distinction to such things as vacuum expectation values whose freedom
can be considered to be part of the initial or boundary conditions). 
Some physicists, such as David Gross, continue to hold out this hope. 
However, it has been discovered that superstring/M theory appears to
have an enormous landscape of possible vacua
\cite{BP,BP2,KKLT,Susskind}, each with different effective constants of
physics (what I have simply called `constants of physics' above and
shall continue to do, since if superstring/M theory is correct, there
are no fundamental true constants of physics other than what can in
principle be deduced from mathematical constants).  Therefore,
superstring/M theory by itself may not give unique predictions for the
constants of physics.

If the constants of physics turn out to be analogous to cosmological
parameters in that they are determined by initial conditions, it might
seem rather hopeless to try to predict them, unless one can get a
definite theory for the initial conditions.  However, the superstring/M
landscape appears to have the property that there can be transitions
between huge sets of the different vacua, so that perhaps some simple
sets of initial states can lead to fairly definite distributions of
vacua and hence of the sets of constants of physics.  This could then
lead to predictions of the statistical distribution of the sets of
constants of physics.  Nevertheless, this distribution is complicated by
the fact that different vacua are expected to lead to different numbers
or different distributions of observers and observations, so that the
statistical distribution of observations has an observership (or
`anthropic') selection effect that modifies the original distribution of
the sets of constants of physics.  There is the further complication
that the numbers of observations for each vacuum can be infinite,
leading to the necessity of performing some regularization of the
results and the corresponding `measure problem' \cite{measure}.  There
are many competing proposals for solving the measure problem which lead
to different statistical distributions of the sets of constants of
physics, but so far no single proposal is so compelling that it has
become universally accepted.

Here I do not wish to get into this controversial issue but instead
adopt the results of Bousso, Hall, and Nomura (BHN) \cite{BHN}, who have
used a particular proposal for the measure that generally seems to fit
observations, to get statistical predictions for the cosmological
constant and the age of the universe at which one would expect
observers.  In Planck units ($\hbar = c = 4\pi\epsilon_0 = G = 1$), they
predict that both $t_\Lambda$ and $t_{\mathrm obs}$ (as well as the
times of galaxy structure formation and galaxy cooling) should be
roughly $\alpha^2/(m_e^2 m_p)$, where $\alpha = e^2/(4\pi\epsilon_0\hbar
c)$ is the fine structure constant (the square of the charge $e$ of an
proton, in relativistic quantum units or Planck units without needing to
set Newton's constant to unity), $m_e$ is the mass of the electron, and
$m_p$ is the mass of the proton.  When one inserts the values for
$\alpha \approx 1/137.036$, $m_e \approx 4.1855\times 10^{-23}$, and
$m_p \approx 7.6851\times 10^{-20}$ in Planck units, one gets
$\alpha^2/(m_e^2 m_p)\approx 3.9554\times 10^{59}\approx 0.676$ Gyr,
about 5\% of the observed age of the universe, 13.7 Gyr $\approx
8.01\times 10^{60}$ Planck times.  (A simple formula for getting closer
to the age of the universe, though without the theoretical justification
that the BHN one above has, is $(\alpha/m_e)^3 \approx 5.3\times 10^{60}
\approx 9.05$ Gyr, 66\% of the age of the universe, within 1\% of how
old the universe was at the formation of the solar system.)  Although
the BHN formula is too small for the actual age of the universe by a
factor of just over 20, the logarithm of the BHN number is only about
2.2\% smaller than the logarithm of the present age.

The Bousso-Hall-Nomura formula gives a very nice relation between
various times within the universe (which are observed to be similar and
which BHN show can be expected to be given approximately by the same
formula) and the three constants of physics that are important for the
hydrogen atom (and for gravity, in order to get Planck units).  However,
by itself the BHN formula does not predict what those three constants
are, so it alone does not predict the absolute age of the universe. 
Nevertheless, there has been other work giving approximate relations
between those three constants of physics and eventually giving all three
purely in terms of mathematical constants
\cite{Carter,Carter2,Carr-Rees,DNP}.

First, Carter \cite{Carter,Carter2} and Carr and Rees \cite{Carr-Rees}
gave anthropic reasons from stellar and nuclear physics that $m_p^2 \sim
\alpha^{12}(m_e/m_p)^4$ and that $m_e/m_p \sim 10\alpha^2$.  The factor
of 10 was just a crude estimate for a numerical coefficient not
predicted precisely, so let us drop it for simplicity (at the cost of
increasing the error).  Then one gets $m_p \sim \alpha^{10} = e^{20}$
and $m_e \sim \alpha^{12} = e^{24}$, where $e = \sqrt{\alpha}$ shall
always be used for the charge of the proton in this paper, not to be
confused with the base of the natural logarithms.  Using the observed
values, the left hand sides of the approximate relations are larger than
the right hand sides by factors of about 179 and 1836 respectively [the
latter of which, being close to $m_p/m_e$, leads to the accidental
coincidence that $(m_p/m_e^2)^{1/12} \approx 137.04$ agrees with the
inverse of the fine structure constant to these five digits], so the
simplified Carter-Carr-Rees formulas, with the numerical factor of 10
dropped, is not very accurate for the quantities themselves, but the
logarithms are within 12\% and 15\%, respectively, of the correct
logarithms.

Now if we insert the simplified Carter-Carr-Rees formulas for the masses
of the proton and electron in terms of the fine structure constant into
the Bousso-Hall-Nomura formula, we get that the age of the universe (or
of vacuum domination, or of structure formation, or of galaxy cooling)
is very roughly $\alpha^{-32} = e^{-64}$.  Using the observed value of
the fine structure constant, this numerically gives $2.3917\times
10^{68}$ Planck units or $4\times 10^{17}$ years, about $3\times 10^7$
times the actual age of the universe.  This relative error is huge, but
one may not be that surprised after dropping the factor of 10 and
raising the inverse fine structure constant to the 32nd power.  On the
other hand, the error in the logarithm is only about 12\%, no worse than
in the simplified Carter-Carr-Rees formulas for the masses of the proton
and electron purely in terms of the powers of the fine structure
constant that they got by anthropic reasoning but leaving out all
numerical factors.

One can see that if one had retained the fudge factor of 10 in the
Carter-Carr-Rees formulas, they would have given $m_p \sim
100\alpha^{10} = 100e^{20}$ and $m_e \sim 1000\alpha^{12} = 1000e^{24}$,
with the right hand sides now too small by only factors of about 1.79
and 1.84.  These inserted into the Bousso-Hall-Nomura formula gives $t
\sim 10^{-8}\alpha^{-32} \approx 2.3917\times 10^{60}\approx 4$ Gyr,
only a factor of three or so too small.  So the error of the simplified
Carter-Carr-Rees formulas is certainly within the slop of (mostly
unknown) numerical factors dropped in the anthropic arguments.

The Bousso-Hall-Nomura formula for the cosmological constant, in Planck
units, and dropping numerical factors that distinguish $\Lambda$ from
the corresponding energy density $\rho_\Lambda = \Lambda/(8\pi)$, is
$\Lambda \sim 1/t_\Lambda^2 \sim m_e^4 m_p^2/\alpha^4$.  Then when one
inserts the simplified Carter-Carr-Rees formulas, one gets $\Lambda \sim
\alpha^{64} = e^{128}$.  The observed value for the fine structure
constant gives $1.75\times 10^{-137}$ which is about $5\times 10^{-16}$
times the observed value of the cosmological constant, $3.5\times
10^{-122}$ in Planck units, but the logarithm is only off by about
12.5\%.  Retaining the Carter-Carr-Rees fudge factor of 10 gives instead
$\Lambda \sim 10^{16}\alpha^{64}$, which is only 5 times larger than the
observed value of the cosmological constant.

We have now expressed the age of the universe (and of structure
formation, galaxy cooling and vacuum domination) purely in terms of the
fine structure constant, but as yet we have no value predicted for it. 
However, I have given a crude renormalization-group argument \cite{DNP}
that $\alpha^{-1}\approx -(10/\pi)\ln{m_p}$, assuming three generations
of quarks and leptons and two relatively light Higgs doublets of
low-energy $SU(3)\times SU(2)\times U(1)$.  When one combines this with
the simplified Carter-Carr-Rees formula $m_p \sim \alpha^{10}$, one
gets  $\alpha\ln{\alpha} \sim -\pi/100$.  Let us denote the solution to
this equation (taken as an exact equation) as $\alpha_a \approx
0.006175533381 \approx 1/161.9293328$.  The square root of this quantity
is an anthropic estimate for the magnitude of the charge of the proton,
$e_a = \sqrt{\alpha_a} \approx 0.07858$, which is about 0.920 times the
observed value of the magnitude of the charge of the proton in Planck
units, $e = \sqrt{\alpha} \approx 0.08542$.  Therefore, using only
observed integers as input (such as the 10th power in the
Carter-Carr-Rees formula that came from, among other things, the
dimension of space, and the 3 for the number of generations and the 2
for the number of light Higgs doublets), and {\it no} observed
continuous parameters, one can get an estimate for the charge of the
proton that has only an 8\% error.  The error in the charge is only
about 2.25\% if one uses the numerical factor of 10 in the
Carter-Carr-Rees formula, but in these crude anthropic estimates it is
rather dubious as to whether this refinement is justified before
comparison with observations.

Now we can insert this mathematically-determined estimate $\alpha_a$ for
the fine structure constant into the Bousso-Hall-Nomura formulas for the
age of the universe and the cosmological constant to get crude predicted
values for them that do not use as input any observation of any
continuous (e.g., non-integer) parameter.  Doing this gives an estimate
for the age of the universe when observers are present (and also for
structure formation, galaxy cooling, and vacuum domination) that is $t_a
\sim \alpha_a^{-32} = e_a^{-64} \approx 5\times 10^{70}$.  Although this
is over six billion times the observed age of about $8\times 10^{60}$ in
Planck units, the logarithm of it is only about 16\% larger than the
logarithm of the observed age.  Similarly, one gets an an anthropic
estimate for the cosmological constant that is $\Lambda \sim
\alpha_a^{64} = e_a^{128} \sim 4\times 10^{-142}$, which is about
$10^{-20}$ times the observed value of $3.5\times 10^{-122}$, but the
predicted logarithm is again only about 16\% larger in magnitude than
the logarithm of the observed value in Planck units.

One can go on to get other quantities like the anthropically predicted
four-volume in the universe to the past of an observer, which is crudely
$V_4 \sim t_a^4 \sim \alpha_a^{-128} = e_a^{-256} \sim 6\times
10^{282}$, more than forty orders of magnitude larger than what our
universe has, largely because of the magnification of the errors by the
huge powers, but with a logarithm only about 16\% too large.

Another quantity that may be estimated is the 3-volume of the $t=t_0$
hypersurface of homogeneity out to a distance that corresponds to
comoving geodesics that one can just barely see, i.e., that intersected
our past light cone at matter decoupling, which on the present
hypersurface of homogeneity may be considered to be the observable
universe today (e.g., what we can see extrapolated to today).  (If one
went all the way back through an inflationary period with very many
e-folds, our past light cone would have spread to enormous comoving
size, so one should cut it off before it gets back to inflation if one
wants a reasonable answer.)  If one does this with the the Mnemonic
Universe Model (MUM) \cite{Agnesi}, which is a spatially flat FRLW model
dominated by dust and $\Lambda$ with present age $t_0 = H_0^{-1} = 10^8$
years$/\alpha$ that is consistent with the measurements to date, then
one gets a radius of about 47.7792 Gyr or $3.48662 t_0$ and a 3-volume
$V_3 \approx 177.542 t_0^3 \approx 4.56883\times 10^5$ Gyr$^3 \approx
1.43578\times 10^{55}$ s$^3 \approx 3.86858\times 10^{80}$ m$^3 \approx
9.16275\times 10^{184}$ cubic Planck lengths.  In comparison, the
anthropic estimate is $t_a^3 = \alpha_a^{-96} = e_a^{-192} \approx
1.2\times 10^{212}$ Planck units, which is over $10^{27}$ times as
large, though the logarithm is only 15\% too large.

Instead of giving the volume of the observable universe today in
astronomical, SI, and Planck units, one might give it in units of the
volume of an observer, say of a 70 kg human, with a volume of about 0.07
m$^3$.  The value given above for the MUM is greater by a factor of
about $5.5\times 10^{81}$.  For an anthropic estimate, one may use an
estimate of the tallest running, breathing organism on a habitable
planet \cite{giraffe}, which gives a height of $L \sim
\alpha^{-0.7}m_e^{-0.95}m_p^{-0.65} \approx 1.512\times 10^{35} \approx
2.44$ meters.  If one said that the 3-volume were $L^3 \approx
3.454\times 10^{105}$ cubic Planck units and divided this into the cube
of the Bousso-Hall-Nomura formula for the age of the universe in Planck
units using the observed values of the fine structure constant and the
mass of the electron and proton, $6.188\times 10^{178}$ cubic Planck
volumes, one would get $1.79\times 10^{73}$, which is smaller than the
ratio for the MUM and the 70 kg human by a factor of $3\times 10^8$,
though the logarithm is only smaller by about 10\%. One can also use
purely the mathematical formulas with $\alpha$ being replaced by the
anthropic estimate $\alpha_a$ that solves $\alpha_a\ln{\alpha_a} =
-0.01\pi$ and with $m_p$ replaced by $\alpha_a^{10}$ and with $m_e$
replaced by $\alpha_a^{12}$.  This then gives the height of the tallest
organism as $L_a \sim \alpha_a^{-18.6}$ and a volume of $L_a^3 \sim
\alpha_a^{-55.8}$.  In units of this anthropic estimate for the volume
of an observer, the anthropic estimate of the 3-volume of the universe
is $(t_a/L_a)^3 \sim \alpha_a^{-40.2} \approx 6.5\times 10^{88}$, which
is a factor of eleven million larger than the observational ratio of
about $5.5\times 10^{81}$ for the 3-volume of the present observed
universe to the volume of a 70 kg human, but the logarithm is only about
9\% larger.

Thus we see that using anthropic arguments in astrophysics and in the
multiverse and combining them with a renormalization group argument, one
can give anthropic estimates for the size of the observable universe and
the magnitude of the cosmological constant purely in terms of
mathematical quantities defined without the input of any non-integer
observed quantities, though the arguments do use observed integers such
as the dimension of space and the number of generations of quarks and
leptons.  Because these quantities are so far from unity in Planck units
(up to about 240 orders of magnitude or so for the 4-volume within our
past light cone), it is not surprising that the estimates are often many
orders of magnitude (up to 40 orders or so for the 4-volume), but the
errors on the logarithms are usually less than one-sixth the magnitudes
of the logarithms themselves.  Therefore, on a logarithmic scale, the
anthropic estimates are remarkably accurate.

Another proposal \cite{BH,BFLR1,BFRL2} is that the huge size of the
universe in Planck units, and the tiny value of the cosmological
constant, is related to the number of vacua in the landscape.  This
might be so, but the absolute value of the common logarithm of the
cosmological constant is around 120, whereas the usual number given as
an estimate for the common logarithm of the number of vacua is 500
(which might itself be an underestimate by a large factor).  So at
present it appears that the logarithm of the number of vacua may be more
than four times the logarithm of the inverse cosmological constant, many
times the error in the estimates of the cosmological constant from the
considerations given in this paper.  Of course, the arguments used in
this paper are more complex and depend more specifically upon the
observed structure of the effective laws of physics in our part of the
landscape than the more generic arguments of the competing proposal, so
one might hope that such a simpler explanation would be viable. 
However, the success of the present arguments for giving good estimates
for the logarithms of the size of the universe suggests that there may
be strong observer selection effects toward universes that have
something like the anthropic relations that Carter, Carr, and Rees have
discovered, along with the renormalization group properties I have found
that allows one to convert those anthropic relations to definite
predictions of the size of the observable universe.

I have benefited from discussions on this subject with Raphael Bousso,
Juan Maldacena, and others.  I am grateful for the hospitality of the
Perimeter Institute for Theoretical Physics in Waterloo, Ontario,
Canada, where this paper was written up.  This work was supported in
part by the Natural Sciences and Engineering Research Council of Canada.


\end{document}